\documentclass{article}
\usepackage{graphicx}
\usepackage{arxiv}
\usepackage{authblk}

\usepackage{amsmath,amssymb,amsfonts,amsthm,mathtools}
\usepackage{mathrsfs}%
\usepackage[utf8]{inputenc}
\usepackage{hyperref}
\usepackage{enumitem}

\usepackage[round]{natbib}  % \citet / \citep。数字の形式(IEEE など)に戻すときは [numbers,square] と unsrtnat
\usepackage{xcolor}%
\usepackage{textcomp}%

\usepackage{multirow}
\usepackage{subfig}
\usepackage{array}
\usepackage{booktabs}
\usepackage{threeparttable}
\usepackage{colortbl}

\newcolumntype{P}[1]{>{\centering\arraybackslash}m{#1}}

\newcommand{\WRONG}{\textbf{W}}
\newcommand{\pow}[1]{2^{#1}}
\newcommand{\ghissue}[1]{Issue~\href{https://github.com/pytorch/pytorch/issues/#1}{\##1}}  % PyTorch の Issue は参考文献ではなく本文中の番号 + リンクで引く
\newcommand{\ghpr}[1]{\href{https://github.com/pytorch/pytorch/pull/#1}{\##1}}

\begin{document}

\newcommand{\myPaperShortTitle}{Silent Failures Beyond the 32-Bit Index Range}
\newcommand{\myPaperTitle}{
\myPaperShortTitle: \\A Differential Characterization of Large-Tensor Matrix Multiplication in PyTorch's MPS Backend
} % Appleを入れたらpreprintのタイトルが4行になってしまい、さすがに長いので削った
\title{\myPaperTitle}
\date{}

%\author{Junichiro Niimi}

\renewcommand\Authfont{\bfseries}
\setlength{\affilsep}{0em}
% box is needed for correct spacing with authblk
\newbox{\orcid}\sbox{\orcid}{\includegraphics[scale=0.06]{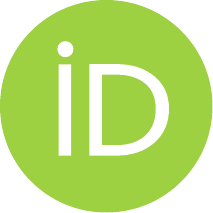}} 
\author[]{%
	\href{https://orcid.org/0000-0002-4618-6272}{\usebox{\orcid}\hspace{1mm}
	Junichiro Niimi\thanks{\texttt{jniimi@meijo-u.ac.jp}}
	}}
\affil[]{Meijo University}

\renewcommand{\shorttitle}{\myPaperShortTitle}

\maketitle

\begin{abstract}
Apple Silicon machines with large unified memory make it possible to hold large tensors on a desktop GPU. However, we found that PyTorch's Metal Performance Shaders (MPS) backend silently returns wrong results for batched matrix multiplication with more than $2^{32}$ elements. \texttt{torch.bmm}, including its wrappers \texttt{torch.matmul} and eager attention, returns relative errors above 1 without an exception or a warning in every PyTorch release tested (2.4.1 to 2.14.0). We sweep \texttt{bmm} over dtypes, memory layouts, shapes and batch sizes around $2^{31}$ and $2^{32}$ elements, and judge every result against a float64 computation on the CPU. Three rules account for every outcome on 2.14.0. When the output exceeds $2^{32}$ elements and an operand is a transposed view, the entire output is wrong and equals a computation that ignores that operand's strides. Otherwise, a view with at least $2^{31}$ elements raises an exception, and a contiguous input above $2^{32}$ elements makes exactly the batches beyond that point wrong, equal to a computation whose index wraps at $2^{32}$. A slightly larger problem can thus turn an explicit error into a silent failure. The rules extend to the backward pass, where a correct forward pass can return silently wrong gradients. A second machine with another chip, under two macOS versions, reproduces all 6156 results, including the wrong values, and the same sweeps on an NVIDIA A100 are correct in all 2530 runs. In a public sentiment classifier, one oversized batch corrupts a third of the outputs, which collapse onto one class. All findings come from observable behavior, without access to the backend's closed-source kernels; we release the harness, raw results and a guard that stops any MPS operation touching $2^{32}$ or more elements at \url{https://github.com/jniimi/mps-silent-failures}.
\end{abstract}

\renewcommand\thefootnote{\arabic{footnote}}
\setcounter{footnote}{0}

\newcommand{\bccol}[2]{ \multicolumn{#1}{c}{\bfseries{#2}}}

\section{Introduction}
\subsection{Background}

With unified memory, the CPU and the GPU of a machine share a single pool of memory, so a model and its data can be placed and moved without the fixed capacity of a separate GPU memory. This flexibility matters more as models, contexts and batches grow, and it has made a desktop machine a practical place to build and analyze large models. Apple Silicon machines offer 192\,GB or more of such memory, and a single fp32 tensor with $\pow{32}$ elements (17\,GB) fits with room to spare. Ordinary workloads reach this size: attention over long contexts or large batches, whose score matrices grow with the square of the sequence length~\citep{tay2022efficient}, feature extraction for probing~\citep{hewitt2019structural}, and attention-map analysis~\citep{clark2019what,abnar2020quantifying}. For example, attention scores of shape $(B \cdot H, L, L)$ pass $\pow{32}$ elements at 32 heads, a context of 4096 tokens and a batch of 9.

PyTorch~\citep{paszke2019pytorch} runs on Apple GPUs through its MPS backend. On macOS 14, batched matrix multiplication whose output exceeded $\pow{32}$ elements stopped with an explicit error (\texttt{Tiling of batch matmul for larger than 2**32 entries only available from MacOS15 onwards}). After an upgrade of the same machine to macOS 27.0, the same operation runs to completion but returns wrong values, without any error or warning (\ghissue{197636}),\footnote{Issue and pull request numbers refer to the PyTorch repository, \url{https://github.com/pytorch/pytorch}.} and we find further silent failures nearby, among them gradients that are silently wrong although the forward pass is correct. For this operation, upgrading the operating system turned a hard failure into a silent one.

Failures at this boundary in this backend are not new. Reports of operations that stop once a tensor reaches $\pow{31}$ or $\pow{32}$ elements go back to 2022 (\ghissue{84039}), and an umbrella issue collects them, stating that the project lacks tooling to detect the affected operations and that the machines available to it cannot allocate tensors of this size (\ghissue{149325}). Batched matrix multiplication with an output above $\pow{32}$ elements became executable in 2024, when the computation was split into tiles (pull requests \ghpr{133430} and \ghpr{143095}) through an interface that Apple provides from macOS 15 on,\footnote{The dependence on macOS 15 is stated in the comment that closes \ghissue{84039}. We did not examine the source ourselves.} which is the change that the two behaviors of the same machine reflect. What these reports leave open is where the boundary of a given operation lies, which tensor has to cross it, what the wrong values are, and whether the backward pass is affected.

Silent failures are among the most costly framework bugs since nothing in the pipeline flags them~\citep{tambon2023silent}. In the case study of Section~\ref{sec:case}, overall accuracy drops by only 7 points while recall for two of the three classes falls to almost zero on the affected inputs, and no error, warning or non-finite value appears.

\subsection{Contributions} 

Our contributions are as follows:

\begin{itemize}[leftmargin=*,itemsep=0pt]
\item \textbf{A method for characterizing failures at a size boundary from the outside.} A boundary-targeted differential sweep of \texttt{bmm} on the MPS backend varies the dtype, the memory layout, the shape, the PyTorch release (11 versions) and which tensor crosses a candidate boundary ($\pow{31}$ elements, $\pow{32}$ elements, $\pow{32}$ bytes), and judges every result against a float64 computation on the CPU. Index-encoded inputs let us read off, without access to the implementation, which input batch, or which row and column within a batch, a wrong output element was computed from.
\item \textbf{A characterization of the failures.} Three rules, applied in order, reproduce all 1376 outcomes with random inputs, including series that probe conditions outside the sweep from which the rules were derived: inputs and output above $\pow{32}$ elements together, batch sizes near $2 \cdot \pow{32}$ elements, and small views at storage offsets beyond $\pow{32}$ elements. The wrong values are identified as two specific misreadings of the inputs. The same rules, applied to the two products that define the gradients, reproduce all 592 outcomes of a sweep of the backward pass, in which a correct forward pass can come with silently wrong gradients.
\item \textbf{Evidence that the failures are deterministic and specific to the backend.} Every series is repeated on a second machine with a different chip, under macOS 26.6.2 and again under macOS 27.0; the three configurations agree in all 6156 common runs, down to the value of the largest error, which separates the chip from the macOS version. The same sweeps on CUDA, including the backward pass, are correct throughout. A first-pass suite of 26 test cases on the 11 releases places \texttt{bmm} among other operations on large tensors and records a regression in 2.14.0, where \texttt{torch.arange} stops raising and silently truncates.
\item \textbf{Impact and mitigation.} In a public sentiment classifier, a single oversized batch corrupts a third of the outputs without any error, the affected predictions collapse onto one class, and chunked execution of the same inputs is correct. A drop-in guard stops any MPS operation that would create or read a tensor with $\pow{32}$ or more elements; we report its overhead and evaluate it against every recorded outcome.
\end{itemize}

\section{Related Work}

\paragraph{Bugs in deep learning frameworks.} Empirical studies classify the bugs reported against deep learning frameworks by symptom and root cause~\citep{chen2023understanding}. \citet{tambon2023silent} study \emph{silent} bugs in Keras and TensorFlow, which produce wrong behavior without a crash or an error message, and find that they are hard for users to notice and to trace back to the framework. \citet{wang2022numerical} study numerical bugs in deep learning programs, such as overflow and loss of precision in floating-point computation. The failures we report are silent in the sense of Tambon et al., but they are not rounding problems: the relative error is of order one and appears only above a fixed element count. These studies classify faults after the fact, from issue trackers and repositories. We instead take a single reported issue (\ghissue{197636}) as the starting point for controlled measurements that establish where the failure occurs, under which conditions, and what the wrong values are.

\paragraph{Reported failures at size boundaries in the MPS backend.} The issue tracker of PyTorch records this size range as a known engineering problem rather than an unexplored one. Operations that fail once a tensor reaches $\pow{31}$ or $\pow{32}$ elements, or 4\,GB, have been reported since 2022 (\ghissue{84039}), and an umbrella issue collects them, noting that tooling to detect the affected operations is still needed and that the machines available for testing cannot allocate tensors of this size (\ghissue{149325}). Most of these reports describe an exception or an assertion, but silent failures appear among them as well. A transfer to the device leaves elements at zero once the tensor reaches $\pow{32}$ bytes (\ghissue{96716}); more recently, a strided copy wraps its writes past an element offset of $\pow{32}$ (\ghissue{182052}), indexing returns zero rows above $\pow{31}$ elements (\ghissue{188756}), and fused attention returns wrong values once its score matrix exceeds $\pow{32}$ elements (\ghissue{179352}). Matrix multiplication has been reported to return wrong results as well, at other boundaries: on earlier Apple GPU generations when the reduction dimension reaches $\pow{15}$ (\ghissue{195163}), and, for the backward pass alone while the forward pass is correct, above about $\pow{15}$ elements and depending on what ran before in the same process (\ghissue{177116}). These reports establish the problem class and several of its instances, each as a single case. What they do not provide is a measurement of where the boundary of one operation lies, of which tensor has to cross it, and of what the wrong values are, across memory layouts, dtypes, library versions and the backward pass, which is what this paper contributes.

\paragraph{Testing deep learning libraries.} Without a ground-truth oracle, library testing is usually differential. CRADLE~\citep{pham2019cradle} runs the same model on several Keras backends and flags inconsistent outputs. Fuzzers generate test inputs automatically: FreeFuzz~\citep{wei2022freefuzz} mines API calls from documentation, tests and models, and compares CPU with GPU execution; NNSmith~\citep{liu2023nnsmith} builds valid computation graphs; generation-based differential fuzzing~\citep{liu2023generation} generates whole models and compares libraries on them; TitanFuzz~\citep{deng2023titanfuzz} lets a large language model write the test programs; and $\nabla$Fuzz~\citep{yang2023fuzzing} tests automatic differentiation by comparing outputs and gradients across execution scenarios. The sweep in this paper (Section~\ref{sec:method}) is differential as well, comparing MPS with a float64 result on the CPU, with CUDA only as a control, and its case study (Section~\ref{sec:case}) uses a metamorphic relation~\citep{chen2018metamorphic} between one large batch and its chunks. The difference lies in what is varied. The tools above aim at breadth over APIs and operator combinations and need each test to be cheap, while a single fp32 tensor with $\pow{32}$ elements already takes 17\,GB. To our knowledge, none of them targets element-count boundaries, and none uses the MPS backend as the system under test. We fix a small family of operations and vary size, memory layout and library version instead.

\paragraph{Integer width and size boundaries.} Integer overflow is common even in mature, widely used programs, and developers often misunderstand it~\citep{dietz2015overflow}. \citet{wressnegger2016twice} show that the migration to 64-bit platforms created vulnerabilities that had not existed before: wider integer types and a larger addressable memory let values reach sizes that code written with 32-bit quantities in mind does not handle. Our setting has the same shape, since a machine with 192\,GB of unified memory is what makes a tensor with $\pow{32}$ elements an ordinary object. Closer to numerical software, the count argument of MPI routines is a C \texttt{int}, so that messages with $\pow{31}$ or more elements are hard to express, and a library that adds support for large counts exposed bugs in MPI implementations~\citep{hammond2014intmax}. The candidate boundaries of Section~\ref{sec:method} are of this kind: they are chosen from the widths of integer types, and the test points are placed around them.

\paragraph{Reproducibility across hardware and silent data corruption.} Training results vary between runs and between accelerators because of nondeterministic kernels and differences in floating-point reduction order~\citep{pham2020variance,zhuang2021randomness}. Such variation is small per operation and is expected behavior, yet it can grow over training: \citet{pham2020variance} report that implementation-level nondeterminism alone changes overall accuracy across identical training runs by up to 2.9\%, and per-class accuracy by up to 52.4\%. \citet{zhuang2021randomness} likewise find a far larger effect on per-class metrics and on subgroups of the data than on top-line accuracy. At the other extreme, studies of large fleets report silent data corruption caused by defective processors, which affects individual machines and is often intermittent~\citep{dixit2021sdc,hochschild2021cores}. The failures in this paper belong to neither group. They are deterministic, they reproduce from a few lines of code across 11 PyTorch releases, and in our measurements whether they occur is decided by tensor size, memory layout and library version, so they are software faults rather than noise or hardware defects.

\section{Method}
\label{sec:method}

\paragraph{A black-box study.} PyTorch's MPS backend is open source, but it is built on Apple's Metal Performance Shaders framework, whose kernels are closed source. We therefore treat the backend as a black box: every statement in this paper is derived from inputs that we control and outputs that we observe, judged against a reference that does not involve the GPU. This is the only view of the system that its users have, and everything we measure can be repeated by anyone with an Apple Silicon machine and the public PyTorch wheels. What the method cannot deliver is a root cause; when we say that a wrong output \emph{equals} a particular misreading of the inputs, we state an agreement of values, not a finding about the implementation.

\paragraph{From \texttt{matmul} to \texttt{bmm}.} \texttt{torch.matmul} (the \texttt{@} operator) dispatches on the number of dimensions. For inputs with three or more dimensions it broadcasts the leading batch dimensions, folds them into one, and calls \texttt{torch.bmm} on $(B, n, m)$ and $(B, m, p)$ tensors. The attention mechanism originates in neural machine translation~\citep{bahdanau2015attention}. The form used in current models is the scaled dot-product attention of the Transformer~\citep{vaswani2017attention}, $\mathrm{softmax}(QK^\top/\sqrt{d})V$. The Hugging Face Transformers library~\citep{wolf2020transformers} calls the implementation that evaluates this formula step by step \emph{eager} attention (\texttt{attn\_implementation="eager"})\footnote{\url{https://huggingface.co/docs/transformers/attention_interface}}, in contrast to fused kernels such as FlashAttention~\citep{dao2022flashattention}, which never return the score matrix. Eager attention is therefore the textbook computation and not a special variant. It computes $QK^\top$ and then $\mathrm{softmax}(QK^\top)V$ with \texttt{matmul}, so both products go through \texttt{bmm}, and the scores of shape $(B, H, L, L)$ exist as a tensor. \texttt{nn.Linear} instead flattens its input to two dimensions and calls \texttt{addmm}, and fused \texttt{scaled\_dot\_product\_attention} (SDPA) is a single operator that does not return the score matrix to the caller.

\paragraph{Environment.} The MPS results come from two machines (Table~\ref{tab:env}). Machine A runs every series. Machine B, with a different chip and memory size, repeats every series twice with the same scripts, identified by their SHA-256 hashes in the manifests: first under macOS 26.6.2, and again after the machine was upgraded to macOS 27.0, the version of machine A. Machine A against machine B under macOS 27.0 thus varies the hardware alone, and the two passes of machine B vary the macOS version alone. Results are compared between configurations run by run, in every recorded field except the run identifiers, the description of the environment, the measured times and memory use, and the text that the system writes when a process aborts. PyTorch versions are installed from the official PyPI wheels into fresh environments with \texttt{uv}, one environment per version.

\begin{table}[htb]
\centering
\caption{Test environments. Machines A and B produce the MPS results. Machine C is the hosted runtime (Google Colab) of the CUDA control; its values are those recorded in the manifests, which are identical in all sessions.}
\label{tab:env}
\small
\begin{tabular}{
wl{1.2cm}
wl{4.2cm}wl{4.2cm}wl{5.0cm}
}
\toprule
 & \bccol{1}{machine A} & \bccol{1}{machine B} & \bccol{1}{machine C (CUDA control)} \\
\midrule
Model & Mac Studio (Mac14,14) & MacBook Pro (Mac14,6) & Google Compute Engine VM \\
\cmidrule(lr){1-4}
Chip & Apple M2 Ultra & Apple M2 Max & Intel Xeon @ 2.20\,GHz (12 vCPUs), \\
 & (24-core CPU, 76-core GPU) & (12-core CPU, 38-core GPU) & NVIDIA A100-SXM4-80GB \\
\cmidrule(lr){1-4}
Memory & 192\,GB unified & 96\,GB unified & 167\,GB host, 80\,GB GPU \\
\cmidrule(lr){1-4}
OS & macOS 27.0 (build 26A428) & macOS 26.6.2 (build 25G83), & Ubuntu 24.04.1 LTS, \\
 & & then macOS 27.0 (build 26A428) & driver 580.82.07, CUDA 13.0 \\
\cmidrule(lr){1-4}
Python & 3.12.12 & 3.12.11 & 3.12.3 \\
\cmidrule(lr){1-4}
PyTorch & the 11 versions (2.4.1, 2.5.1,  & the same 11 versions, & 2.14.0+cu130 \\
 & 2.6.0, 2.7.1, 2.8.0, 2.9.1, 2.10.0, & under each macOS version & \\
 &  2.11.0, 2.12.1, 2.13.0, 2.14.0) & & \\
\bottomrule
\end{tabular}
\end{table}

\paragraph{Axes.} The sweep targets one operation, \texttt{torch.bmm} on $a$ of shape $(B, M, K)$ and $b$ of shape $(B, K, N)$, and varies the axes in Table~\ref{tab:axes}. Every shape has $\pow{16}$ elements per batch in the tensor that is meant to cross the boundary (the \emph{crossing tensor}), which is the output ($MN = \pow{16}$) in the output-side shapes and the input $a$ ($MK = \pow{16}$) in the input-side shapes. That tensor therefore has $\pow{16}B$ elements. In the base shapes the other two tensors have $\pow{14}B$ elements each and stay near $\pow{30}$. The additional shapes split the same number of elements into different $(M, K, N)$, to check that the behavior depends on the element count and not on a particular dimension. In each of them one of the other tensors has $\pow{15}B$ elements and reaches $\pow{31}$ at $B = 65536$. The layouts change only how the operands are stored. The logical values of $a$ and $b$ are generated on the CPU from a fixed seed per batch, independently of the layout, and the storage is built from them, so that for the same seed all four layouts compute the same matrix product and their results can be compared directly. Operations other than \texttt{bmm} are covered only by the first-pass suite (Section~\ref{sec:firstpass}).

\begin{table}[t]
\centering
\caption{Axes of the \texttt{bmm} sweeps. In the main sweep, the base shapes are run in both dtypes and with all input kinds, and the additional shapes in fp32 with random inputs. Entries marked \emph{ext.} belong to the extension series and \emph{bwd.} to the backward series, which are described below.}
\label{tab:axes}
\begin{tabular}{lp{0.74\linewidth}}
\toprule
axis & values \\
\midrule
shape $(M, K, N)$ & output-side: base=$(256, 64, 256)$, $(512, 64, 128)$, $(128, 64, 512)$\\
&input-side: ~~base=$(256, 256, 64)$, $(512, 128, 64)$, $(128, 512, 64)$ \\
&inputs and output: $(256, 256, 256)$ (ext.) \\
operation & forward; forward and backward (bwd., base shapes) \\
dtype & fp32, fp16 \\
layout & contiguous; $b$ transposed; $a$ transposed (a contiguous storage of the transposed values, passed as a \texttt{transpose(1, 2)} view); sliced (both operands are \texttt{[1:]} of a storage with one extra leading batch, so the storage offset is one batch); offset (ext.: a view of 4096 batches behind $O$ leading batches, for both operands) \\
input values & random (uniform on $[-1, 1]$, zeros replaced so that every element is non-zero); index-encoded (see below) \\
batch size $B$ & 4096 to 65538 (see below); 131071 to 131073 (ext.: far points, base shapes) \\
PyTorch & 2.14.0 for the main sweep, the extension series and the backward series; 2.4.1 to 2.13.0 for a reduced sweep (base shapes only) \\
\bottomrule
\end{tabular}
\end{table}

\paragraph{Boundaries and search.} We consider three candidate boundaries for the crossing tensor: $\pow{31}$ elements ($B_0 = 32768$), $\pow{32}$ elements ($B_0 = 65536$), and $\pow{32}$ bytes ($B_0 = 16384$ in fp32; in fp16 this coincides with $\pow{31}$ elements). These candidates are hypotheses about where a limit could lie. A 32-bit signed integer can index at most $\pow{31} - 1$ elements and an unsigned one at most $\pow{32} - 1$, so an implementation that holds an element index in either type would change its behavior at the first or the second candidate. The third candidate tests whether a limit applies to the number of bytes instead of the number of elements: in fp32 a tensor reaches $\pow{32}$ bytes at $\pow{30}$ elements, and a byte limit would also make fp16 and fp32 switch at different batch sizes. Each series, that is, each combination of shape, dtype and layout, is run at the five consecutive batch sizes $B_0 - 2, \ldots, B_0 + 2$ around every candidate, at a small baseline point ($B = 4096$, $\pow{28}$ elements), and on an evenly spaced grid between these points (seven values per interval for the base shapes and three for the additional ones). This gives 37 batch sizes per series in fp32 and 25 in fp16 for the base shapes. A change of behavior that does not coincide with a candidate would appear as two neighboring grid points with different outcomes, and the driver then bisects between them down to a single batch. At each switch point, the batch sizes on both sides are run again with a comparison over all elements and three seeds; a series without a switch receives this comparison at its largest batch size.

\paragraph{Reference and comparison.} The reference is the same product computed in float64 on the CPU. To build it, the logical inputs of each batch are generated again on the CPU from their seeds; inputs are never read back from the device for this purpose, so a fault in the transfer cannot cancel out. The device output is copied to the CPU in chunks, compared, and discarded. Let $Y_i$ be batch $i$ of the device output and $Y_i^{\ast}$ the same batch of the reference, and let $\|A\|_{\max}$ denote the largest absolute entry of a matrix $A$. The error of batch $i$ is
\begin{equation}
e_i = \frac{\| Y_i - Y_i^{\ast} \|_{\max}}{\| Y_i^{\ast} \|_{\max}},
\label{eq:err}
\end{equation}
and batch $i$ is wrong when $e_i > \tau_d$, where the tolerance $\tau_d$ is set for each dtype $d$ and each series. Let $\hat{e}_d$ be the largest $e_i$ in a set of calibration runs far below any candidate (every shape of the series that is run in that dtype, $B = 4096$, contiguous, three seeds, all elements). Then
\begin{equation}
\tau_d = \max\bigl( 10\, \hat{e}_d,\; \tau_d^{\min} \bigr),
\label{eq:tol}
\end{equation}
with a floor $\tau_d^{\min}$ of $10^{-6}$ in fp32 and $10^{-3}$ in fp16.

A comparison is either \emph{full}, over all batches, or \emph{sampled}. A sampled comparison takes 256 evenly spaced batches, the first 4 and the last 8, and the batches around every position where the flat index of an operand or of the output passes a multiple of a candidate boundary, which gives 266 to 281 batches. The grid and the points around the candidates use sampled comparisons, and the switch points use full comparisons with three seeds. The reduced sweep for earlier PyTorch versions uses a coarser grid (two values per interval) and one seed for the full comparisons.

Each run is classified as \emph{ok}; \emph{error}, when an exception is raised; \emph{crash}, when the process ends without returning a result; \emph{timeout} (20 minutes for a sampled and 60 for a full comparison); \emph{wrong}; \emph{truncated} (an output of zeros), when at least 90\% of the wrong elements are exactly zero, which the non-zero inputs make detectable; or \emph{input corrupt}, when the inputs read back from the device in the sampled batches differ from the generated values, which would point to the transfer and not to \texttt{bmm}. Runs whose estimated memory exceeds 60\% of the physical memory are skipped. For wrong runs we also test a small set of hypotheses about how the inputs were read. For up to 16 wrong batches at each end of the wrong range, we recompute on the CPU the output that would result if the strides of a transposed operand were ignored, if the flat index into an operand wrapped around at $\pow{32}$ elements, if both happened, or if the storage offset were ignored, and record the fraction of wrong elements that each hypothesis reproduces.

\paragraph{Extensions of the sweep.} Three extension series use the same worker, the same reference and the same classification as the main sweep, and address conditions that it does not reach. (i) \emph{Inputs and output together.} The shape $(M, K, N) = (256, 256, 256)$ gives $a$, $b$ and the output $\pow{16}$ elements per batch each, so that both inputs and the output pass every candidate at the same batch size. It is swept like a base shape, in fp32 and fp16 and in the four layouts. (ii) \emph{Far points.} For the two base shapes, all layouts and both dtypes, we add the batch sizes $2 B_0 - 1$, $2 B_0$ and $2 B_0 + 1$ with $B_0 = 65536$, where the crossing tensor has about $2 \cdot \pow{32}$ elements, with a comparison over all elements and with the four index-encoded inputs at $2 B_0 + 1$. (iii) \emph{Large storage offsets.} To separate the size of a view from its position in memory, a small view of $B = 4096$ batches ($\pow{28}$ elements in the crossing tensor) is placed behind $O$ leading batches in a larger storage, for both operands. Leading batch $j$ is filled with a value that is constant within the batch and differs between batches and from the values of the view, so that a read from the wrong place is detectable. $O$ is swept so that the storage offset of the larger operand passes $\pow{31}$ and $\pow{32}$ elements: $O = 1$, five consecutive values around each candidate, and a grid in between (27 values in fp32, 19 in fp16), all with comparisons over all elements. The hypotheses of wrapped index and ignored offset are adapted to this layout. The tolerance is calibrated separately for each series and is reported with its results.

\paragraph{Backward pass.} To test gradients, the worker runs the forward product $y = \texttt{bmm}(a, b)$ with both operands requiring gradients and then backpropagates an upstream gradient $G$ of the shape of $y$. We write $A_i$, $B_i$, $Y_i$ and $G_i$ for batch $i$ of $a$, $b$, $y$ and $G$, and $\ell$ for the scalar that is differentiated. $G$ is generated on the CPU from its own seeds in the same way as an operand and is contiguous on the device. The references are computed in float64 on the CPU from the regenerated inputs, batch by batch,
\begin{equation}
Y_i^{\ast} = A_i B_i, \qquad \Bigl(\frac{\partial \ell}{\partial a}\Bigr)_i^{\ast} = G_i B_i^{\top}, \qquad \Bigl(\frac{\partial \ell}{\partial b}\Bigr)_i^{\ast} = A_i^{\top} G_i ,
\label{eq:bwd}
\end{equation}
and the output and the two gradients are each compared with the error of Eq.~\eqref{eq:err} and the same tolerance, which is calibrated on the largest error of the three tensors in the calibration runs of this series. A run is classified by the worst of the three tensors, and the class, the largest error and the wrong batches are also recorded per tensor; for an exception we record whether it was raised in the forward or in the backward pass. The layout applies to $a$ and $b$. This series uses the two base shapes, both dtypes, the four layouts and the batch sizes of the main sweep.

\paragraph{CUDA control.} The main sweep, the backward series and the three extension series are repeated on an NVIDIA A100 (machine C in Table~\ref{tab:env}) with the same driver and worker scripts; the worker selects the device with a flag, so that the input generation, the reference, the error of Eq.~\eqref{eq:err} and the classification are identical; the SHA-256 hashes of both scripts are recorded in every manifest and equal those of the MPS runs. TF32 is disabled. Because a hosted session does not last for the whole sweep, each series is split, where necessary, into chunks of one shape, one dtype and one or two layouts (the far points by shape and dtype), each chunk is one run, and only chunks that finished are used. Every chunk is calibrated on its own shape and dtype with Eq.~\eqref{eq:tol}, so the tolerance differs between chunks, and for the control we also report the largest error over all runs. The batch sizes are the same 42 values. On CUDA no switch point exists by construction of the search, so the full comparisons fall on the largest batch size of every series. These runs serve as a control and not as an oracle.

\paragraph{Index-encoded inputs.} Random inputs show that an output is wrong, but not which input elements it was computed from. Index-encoded inputs make every output element carry that information. The idea is long established in memory testing, where the data written to a memory are chosen so that a given class of faults becomes observable from outside~\citep{vandegoor1993march}. One operand is \emph{encoded}: each of its elements holds the code
\begin{equation}
c(f) = (f \bmod P) + 1
\label{eq:code}
\end{equation}
of a field $f$, which is either the batch index $i$ (the same code in the whole batch) or the position $rC + c$ of element $(r, c)$ within its $R \times C$ matrix (the same codes in every batch). The other operand is a \emph{selection} matrix with a single one in each row or column, identical in every batch. When $b$ is encoded, $a_{i,m,k} = [k = m \bmod K]$, and when $a$ is encoded, $b_{i,k,n} = [k = n \bmod K]$, so that a correct product is
\begin{equation}
y_{i,m,n} = b_{i,\, m \bmod K,\, n} \qquad \text{or} \qquad y_{i,m,n} = a_{i,\, m,\, n \bmod K},
\label{eq:select}
\end{equation}
respectively. Each output element is then a sum with one non-zero term and equals one code exactly, without rounding. The period $P$ is the largest prime below $\pow{20}$ in fp32 ($P = 1048573$) and below $\pow{11}$ in fp16 ($P = 2039$), so every code is an integer that the dtype represents exactly. From a wrong output element $y$ we decode the field that was actually read, $f' = y - 1$, and record the shift
\begin{equation}
\delta = f' - f \pmod{P}, \qquad -P/2 \le \delta < P/2,
\label{eq:delta}
\end{equation}
against the field $f$ that a correct product would show at that place. In fp32 the decoding is unique, because both the batch sizes and the $\pow{16}$ positions within a matrix are below $P$; in fp16 it is known only modulo 2039. With the position encoded, $f'$ gives the row and column that were read in place of the expected ones. The two operands and the two fields give four kinds of input, which we run for the base shapes at $B_0 \pm 1$ around every candidate and at $B = 4096$ with sampled comparisons, and at the switch points with full comparisons.

\paragraph{Isolation.} Each run executes in a fresh process, because a failure may abort the process, and a wrong result may corrupt state for later calls in the same process. A crash is recorded as an outcome. The driver appends one JSON line per run to a local file, with the configuration, the environment, the outcome, the error statistics and the timings. A manifest per sweep records the operating system build, the hardware, the Python and PyTorch versions, the tolerances and the SHA-256 hashes of the driver and the worker script. These files are released with the code.

\section{Results}

\subsection{Boundary sweep of \texttt{bmm}}
\label{sec:sweep}

The main sweep on PyTorch 2.14.0 comprises 1596 runs over 42 batch sizes between 4096 and 65538, 24 of them calibration runs: 1285 are correct, 214 raise an error, and 97 are silently wrong. Here and below, the number of runs of a series includes its calibration runs, which are all correct; the comparisons with the rules and between machines exclude them, as stated there. No run crashed or timed out, and no wrong output contained a non-finite value. Sampled and full comparisons agree wherever both were run, and so do the three seeds. The largest calibration errors are $\hat{e}_{\mathrm{fp32}} = 1.5\times10^{-6}$ and $\hat{e}_{\mathrm{fp16}} = 4.2\times10^{-4}$, so the tolerances of Eq.~\eqref{eq:tol} are $\tau_{\mathrm{fp32}} = 1.5\times10^{-5}$ and $\tau_{\mathrm{fp16}} = 4.2\times10^{-3}$.

\begin{figure}[htb]
\centering
\includegraphics[width=\linewidth]{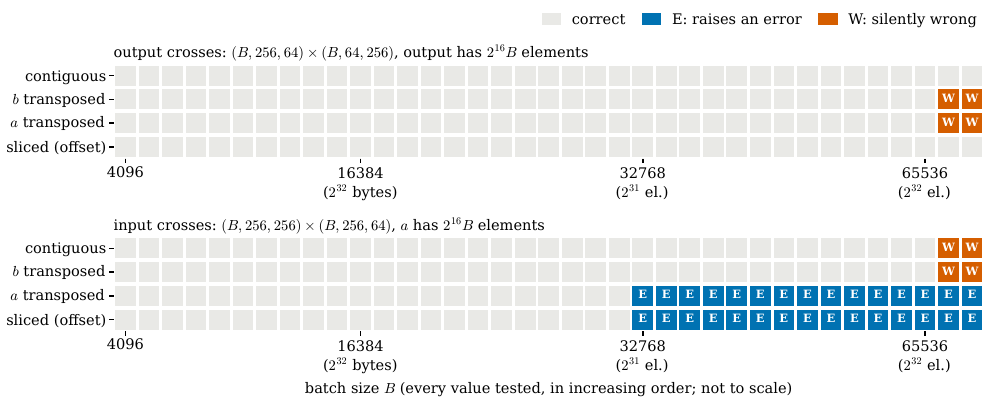}
\caption{Boundary map of \texttt{bmm} on MPS (PyTorch 2.14.0, macOS 27.0, fp32, random inputs). Each cell is one batch size; five consecutive values are tested around each candidate boundary, and the remaining values form a grid. The fp16 results have the same pattern at the same batch sizes.}
\label{fig:map}
\end{figure}

\paragraph{Boundary map.} Figure~\ref{fig:map} shows the two base shapes. Nothing changes at $\pow{32}$ bytes. The silent failures start between $B = 65536$ and $B = 65537$, that is, the result is still correct at exactly $\pow{32}$ elements and wrong one batch later. fp16 switches at the same batch sizes as fp32, so the relevant quantity is the number of elements and not the number of bytes. The exceptions start at $B = 32768$, where the view operand has exactly $\pow{31}$ elements, and carry the message \texttt{MPSGraph does not support tensor dims larger than INT\_MAX}. Every change of behavior lies between two consecutive batch sizes next to a candidate, so no bisection was triggered and the switch points are exact.

The outcome of every run with random inputs (1076 runs outside calibration, six shapes, two dtypes, four layouts) is reproduced by the rules in Table~\ref{tab:rules}, applied in order. The order matters. With $a$ of shape $(B, 512, 64)$ passed as a transposed view, the operation is correct at $B = 65535$, raises at $B = 65536$, where $a$ reaches $\pow{31}$ elements while the output has exactly $\pow{32}$, and is silently wrong at $B = 65537$, where the output exceeds $\pow{32}$. With sliced operands of the same shape it raises at $B = 65536$ and is correct again at $B = 65537$. A larger problem can therefore turn an explicit error into a silent failure. The rules describe what we observed on this PyTorch version and are not a specification. No configuration of the main sweep has an output above $\pow{32}$ elements together with an input above $\pow{32}$, so it cannot tell whether the first rule also takes precedence over the third; Section~\ref{sec:extended} measures this case.

\begin{table}[htb]
\centering
\caption{Rules that reproduce all 1076 outcomes of the sweep with random inputs (PyTorch 2.14.0), applied in order. A \emph{view} is an operand passed as a transposed or a sliced tensor.}
\label{tab:rules}
\small
\begin{tabular}{clll}
\toprule
& condition & outcome & extent \\
\midrule
1 & output $> \pow{32}$ elements, an operand is transposed & silently wrong & every batch of the output \\
  & output $> \pow{32}$ elements, otherwise & correct & \\
2 & an operand that is a view has $\geq \pow{31}$ elements & raises an error & \\
3 & a contiguous operand has $> \pow{32}$ elements & silently wrong & only batches beyond element $\pow{32}$ \\
 & otherwise & correct & \\
\bottomrule
\end{tabular}
\end{table}

\paragraph{Two kinds of wrong result.} The two silent failures differ in extent. When the output crosses the boundary and an operand is transposed (rule 1), all 65537 batches are wrong, including batch 0, and the largest error $e_i$ of Eq.~\eqref{eq:err} is between 1.7 and 2.2. Exceeding the boundary by a single batch thus invalidates the whole output. When a contiguous input crosses the boundary (rule 3), only the batches that lie beyond element $\pow{32}$ of that input are wrong, with $e_i$ between 1.4 and 1.7, and all earlier batches pass the comparison. This is the pattern of the case study in Section~\ref{sec:case}, where the texts before position 1365 are unaffected.

\paragraph{What the wrong values are.} For each wrong run we tested the four hypotheses of Section~\ref{sec:method} about how the inputs were read, and counted the wrong elements that each reproduces. Under rule 1, every wrong element equals the value obtained when the strides of the transposed operand are ignored, that is, when its storage is read as if it were contiguous. Under rule 3, every wrong element equals the value obtained when the flat index into the large operand wraps around at $\pow{32}$ elements, so that batch 65536 is computed from batch 0. The index-encoded inputs confirm both readings element by element without reference to a hypothesis. With the batch index encoded in $a$, the shift of Eq.~\eqref{eq:delta} is $\delta = -65536$ for every wrong element: the decoded batch lies exactly $\pow{16}$ batches, that is $\pow{32}$ elements, before the expected one. In fp16 the shift is $\delta = -288$, which is $-65536$ modulo $P = 2039$. With the position inside the matrix encoded in a transposed $b$ of logical shape $64 \times 256$, the output that should contain element $(0, 1)$ contains element $(1, 0)$, and the one that should contain $(16, 0)$ contains $(0, 64)$, as expected when a $256 \times 64$ storage is read row by row as $64 \times 256$. The three rules predict the outcome of 460 of the 496 runs with index-encoded inputs. The other 36 are predicted to be wrong but are correct, and they are explained by the inputs themselves. Some index-encoded inputs do not change under the misreading, for example values that are constant within a batch when strides are ignored, or operands that are identical in every batch when the index wraps by whole batches. We checked this on the CPU for all 64 index-encoded runs in which the rules predict a wrong result, by computing the product under the predicted misreading. It equals the correct product exactly in the 36 runs that were recorded as correct, and differs from it in the 28 runs that were recorded as wrong. With this taken into account, the rules account for all 1572 runs outside calibration.

\paragraph{Earlier PyTorch versions.} Table~\ref{tab:sweepversions} summarizes the reduced sweep of the base shapes on ten earlier releases ($\tau_{\mathrm{fp32}} = 1.2\times10^{-5}$ and $\tau_{\mathrm{fp16}} = 4.2\times10^{-3}$ in every release). The releases fall into four groups with identical outcomes within each group. Two results hold in every release since 2.5.1: a transposed operand with an output above $\pow{32}$ elements gives a wrong output that matches the ignored-strides computation, and a contiguous input above $\pow{32}$ elements gives wrong batches that match the wrapped index. What changed over time is how an oversized \emph{view} is handled. In 2.4.1 it is accepted below $\pow{32}$ elements and returns an output of zeros from $\pow{32}$ elements on (the truncated outcome; 16 runs, all on 2.4.1). From 2.5.1 to 2.8.0 it aborts the process with a Metal assertion (\texttt{NDArray dimension length > INT\_MAX}) between $\pow{31}$ and $\pow{32}$ elements, raises an exception at exactly $\pow{32}$ elements, and above $\pow{32}$ elements is silently wrong with the wrapped index. From 2.9.1 on it raises an exception everywhere from $\pow{31}$ elements upward. For a sliced operand the thresholds up to 2.9.1 apply to the storage, which holds one batch more than the view, so they are reached one batch earlier. Version 2.4.1 differs in two further ways. With an fp32 output above $\pow{32}$ elements it is wrong for every layout, including contiguous operands, but only in the first and the last batch, and half of the wrong elements are zero; none of our hypotheses reproduces these values. With fp16 the same configurations are correct. Across all versions no wrong output contains a non-finite value, no run is classified as input corrupt or timeout, no bisection was triggered, and within 2.10.0 to 2.14.0 we found no difference for \texttt{bmm}.

\begin{table}[htb]
\centering
\caption{Outcome of \texttt{bmm} by PyTorch version (base shapes, fp32 and fp16 unless noted, macOS 27.0). $n$ is the number of elements of the tensor in question. \WRONG: silently wrong; Z: output of zeros, silently; ok: correct; err: raises; crash: process aborts.}
\label{tab:sweepversions}
\small
\begin{tabular}{lcccc}
\toprule
condition & 2.4.1 & 2.5.1--2.8.0 & 2.9.1 & 2.10.0--2.14.0 \\
\midrule
output $n > \pow{32}$, contiguous or sliced & \WRONG\ (fp32), ok (fp16) & ok & ok & ok \\
output $n > \pow{32}$, an operand transposed & \WRONG\ (fp32), ok (fp16) & \WRONG & \WRONG & \WRONG \\
contiguous input $n > \pow{32}$ & \WRONG & \WRONG & \WRONG & \WRONG \\
view input, $\pow{31} \le n < \pow{32}$ & ok & crash & err & err \\
view input, $n = \pow{32}$ & Z & err & err & err \\
view input, $n > \pow{32}$ & Z & \WRONG & err & err \\
\bottomrule
\end{tabular}
\end{table}

\subsection{Extension series}
\label{sec:extended}

\paragraph{Inputs and output above $\pow{32}$ together.} With $(M, K, N) = (256, 256, 256)$, fp32 and fp16 behave identically (228 runs, and 36 further runs that repeat the wrong and the neighboring correct points with comparisons over all elements and three seeds). Contiguous operands are correct at every batch size up to $B = 65538$, although $a$ and $b$ then exceed $\pow{32}$ elements; the largest error is $1.3\times10^{-6}$ in fp32 and $4.8\times10^{-4}$ in fp16. With a transposed operand the operation is correct up to $B = 32767$, raises the \texttt{INT\_MAX} exception from $B = 32768$ to $65536$, where the view has at least $\pow{31}$ elements, and is silently wrong in every batch from $B = 65537$, where the output exceeds $\pow{32}$ elements (24 comparisons over all elements, largest error between 2.05 and 2.19, all wrong elements reproduced by the ignored strides). With sliced operands it raises over the same range and is correct again from $B = 65537$. This is exactly what Table~\ref{tab:rules} predicts: once the output exceeds $\pow{32}$ elements, the first rule decides the outcome, and neither the exception for large views nor the wrapped index of a large contiguous input occurs.

\paragraph{About $2 \cdot \pow{32}$ elements.} At $B = 131071$, $131072$ and $131073$ the picture does not change (124 runs, all elements compared, fp32 and fp16 identical). With the output-side shape, contiguous and sliced operands are correct and a transposed operand makes every batch wrong (largest error 2.16 to 2.18). With the input-side shape and contiguous operands, the wrong batches are exactly those from 65536 to $B - 1$, that is, every batch beyond element $\pow{32}$ of $a$, and every wrong element matches the index wrapped at $\pow{32}$; at $B = 131073$ this includes batch 131072, which lies beyond $2 \cdot \pow{32}$ elements, so the index wraps a second time. A transposed $b$ of shape $(B, 256, 64)$ gives the same wrong batches at $B = 131071$ and raises from $B = 131072$, where this view reaches $\pow{31}$ elements; the second rule thus applies to either operand. A transposed or sliced $a$ raises at all three batch sizes. Of the 24 runs with index-encoded inputs for which the rules predict a wrong result, 12 are wrong and 12 are correct, and the check of Section~\ref{sec:sweep} shows that the 12 correct ones are those whose input does not change under the misreading.

\paragraph{Large storage offsets.} All 104 runs of the offset series are correct in every element, in fp32 and fp16 and for both base shapes. This includes views that straddle element $\pow{31}$ or $\pow{32}$ of their storage and views that lie entirely beyond it; the largest storage offset is $\pow{32} + \pow{17}$ elements in a storage of $4.56\times10^{9}$ elements. A small contiguous view is therefore computed correctly wherever it lies in a large device tensor, on PyTorch 2.14.0. We did not combine a large offset with a transposed view.

The tolerances of these series are $\tau_{\mathrm{fp32}} = 1.15\times10^{-5}$ and $\tau_{\mathrm{fp16}} = 3.9\times10^{-3}$ for the $(256, 256, 256)$ shape, and $1.23\times10^{-5}$ and $4.2\times10^{-3}$ for the far points and the offset series; every wrong result exceeds them by five orders of magnitude or more in fp32.

\paragraph{CUDA control.} On the A100, the three extension series are correct in every run: 204 runs of the $(256, 256, 256)$ shape and 36 runs of its repeated points, 124 runs at the far points, and 104 runs of the offset series, with a largest error of $1.4\times10^{-6}$ in fp32 and $4.8\times10^{-4}$ in fp16.

\paragraph{The rules against all runs.} Table~\ref{tab:rulecheck} applies the rules of Table~\ref{tab:rules}, unchanged, to every run with random inputs. They reproduce all 1376 outcomes. The case that the main sweep could not decide is settled by the 16 runs with contiguous operands of the $(256, 256, 256)$ shape above $\pow{32}$ elements: a reading in which the third rule still applied there would predict wrong batches, and all 16 are correct.

\begin{table}[htb]
\centering
\caption{The rules of Table~\ref{tab:rules} applied to every run with random inputs (PyTorch 2.14.0, macOS 27.0; calibration runs excluded). For wrong runs with a comparison over all elements, agreement includes the extent: every batch, or exactly the batches beyond element $\pow{32}$.}
\label{tab:rulecheck}
\small
\begin{tabular}{lrr}
\toprule
series & runs & outcome reproduced \\
\midrule
main sweep (Section~\ref{sec:sweep}) & 1076 & 1076 \\
$(256, 256, 256)$ shape, sweep & 222 & 222 \\
$(256, 256, 256)$ shape, repeated points & 30 & 30 \\
far points & 48 & 48 \\
\midrule
all & 1376 & 1376 \\
\bottomrule
\end{tabular}
\end{table}

\subsection{Backward pass}
\label{sec:backward}

The backward series comprises 604 runs on PyTorch 2.14.0 ($\tau_{\mathrm{fp32}} = 1.28\times10^{-5}$, $\tau_{\mathrm{fp16}} = 4.2\times10^{-3}$): 420 are correct in all three tensors, 40 are silently wrong in at least one, and 144 raise an exception. fp32 and fp16 agree at every point.

\paragraph{Correct forward pass, wrong gradients.} With the output-side shape, everything is correct up to $B = 65536$. At $B = 65537$ and $65538$, with contiguous or sliced operands, the forward output is correct, as in Section~\ref{sec:sweep}, but both gradients are silently wrong (20 runs). The wrong batches of $\partial\ell/\partial a$ and $\partial\ell/\partial b$ are exactly those from 65536 on, that is, the batches beyond element $\pow{32}$ of $G$; the comparisons over all elements at $B = 65537$ with three seeds show a single wrong batch, the last one, in every layout and dtype. The largest error is between 1.16 and 1.57, and every wrong element matches the index wrapped at $\pow{32}$. With a transposed operand, the forward output is wrong in every batch, as before, and the gradients are wrong in the same last batches (20 runs). A model whose forward pass is correct at this size can therefore still be trained on wrong gradients, without an error or a non-finite value.

\paragraph{Exceptions.} With the input-side shape, everything is correct up to $B = 32767$, and every run from $B = 32768$ raises the \texttt{INT\_MAX} exception. For a transposed or sliced $a$ it is raised in the forward pass, as in Section~\ref{sec:sweep}. For contiguous operands and for a transposed $b$, the forward pass succeeds and the exception is raised in the backward pass (72 of the 144 runs). At these sizes the failure is loud.

\paragraph{The same rules.} These outcomes are what Table~\ref{tab:rules} gives when it is applied to the two products of Eq.~\eqref{eq:bwd}, with $G$ as a contiguous operand and each transpose as a view: the transpose of a contiguous operand is a transposed view, and the transpose of an operand that was passed as a transposed view is its contiguous storage. For the output-side shape, $G$ is a contiguous operand with more than $\pow{32}$ elements from $B = 65537$ while the gradients themselves stay below $\pow{32}$ elements, which is the third rule. For the input-side shape, the transpose of $a$ is a view with $\pow{31}$ elements at $B = 32768$, which is the second rule. Applied in this way, the rules reproduce the class of the output and of both gradients in all 448 runs that return a result, including the extent of the wrong batches where all elements were compared, and the pass in which the exception occurs in all 144 runs that raise. We state this as an agreement between outcomes and rules; how the backward pass forms its operands was not observed.

\paragraph{CUDA control.} On the A100, all 568 runs of the backward series are correct in all three tensors (24 of them calibration runs). This includes the batch sizes $B = 65537$ and $65538$ of the output-side shape, where MPS returns wrong gradients, and the runs of the input-side shape from $B = 32768$, which raise on MPS. The largest error is $1.5\times10^{-6}$, $1.4\times10^{-6}$ and $1.4\times10^{-6}$ for the output and the two gradients in fp32, and at most $4.9\times10^{-4}$ for each of them in fp16. The series has fewer runs than on MPS (604) because there are no switch points to compare on both sides.

\subsection{A second machine and two macOS versions}
\label{sec:machines}

Machine B repeats all 6348 runs of machine A under each of its two macOS versions, 6156 of them outside calibration. Table~\ref{tab:machines} compares the three configurations run by run, where a run is identified by its operation, shape, dtype, layout, input, batch size, seed and kind of comparison, and in the offset series also by the number of leading batches. In each of the three pairs, every run of one configuration has its counterpart in the other, and in all 6156 the two agree in every compared field: the class, the largest error of Eq.~\eqref{eq:err} as a floating-point value, the ranges of wrong batches, the number of wrong elements, the fractions of wrong elements that each hypothesis reproduces, the shifts decoded from index-encoded inputs, the message of each exception, and, for the backward series, the same quantities per tensor and the pass in which an exception is raised. The calibration errors, and hence the tolerances, are identical as well. The agreement covers the wrong results as much as the correct ones. It includes the behavior that is specific to old PyTorch versions: the outputs of zeros and the wrong values of 2.4.1 that no hypothesis reproduces, and the process aborts of 2.5.1 to 2.8.0. The comparison is produced by a script that is released with the raw results and fails on any difference.

\begin{table}[htb]
\centering
\caption{Machine A (M2 Ultra, macOS 27.0) and machine B (M2 Max) under macOS 26.6.2 and under macOS 27.0, compared run by run, calibration runs excluded. The counts are the same in all three pairs of configurations. ``Identical'' means equal in every compared field, among them the class, the largest error, the wrong batches, the hypothesis fractions and the tolerance.}
\label{tab:machines}
\small
\begin{tabular}{llrrr}
\toprule
series & PyTorch & common runs & identical & only in one configuration \\
\midrule
main sweep & 2.14.0 & 1572 & 1572 & 0 \\
reduced sweep & 2.4.1, 2.9.1--2.13.0 (each) & 348 & 348 & 0 \\
reduced sweep & 2.5.1--2.8.0 (each) & 362 & 362 & 0 \\
inputs and output together & 2.14.0 & 252 & 252 & 0 \\
far points & 2.14.0 & 112 & 112 & 0 \\
large storage offsets & 2.14.0 & 92 & 92 & 0 \\
backward pass & 2.14.0 & 592 & 592 & 0 \\
\midrule
all & & 6156 & 6156 & 0 \\
\bottomrule
\end{tabular}
\end{table}

The failures are therefore not a property of one machine, of the M2 Ultra, or of one macOS version, and they are deterministic to the last bit of the reported error. The three configurations separate the two factors. Machine A and machine B under macOS 27.0 run the same build of the operating system on different chips, and the two passes of machine B run different macOS versions on the same hardware; neither changes any result. The main sweep takes 116 and 118 minutes on machine B and 114 minutes on machine A, although machine B has half the GPU cores, because the run time is dominated by generating the inputs and computing the float64 reference on the CPU.

\subsection{Case study: sentiment classification with eager attention}
\label{sec:case}

To check that the failure reaches real workloads, we ran a public sentiment classifier through the standard Hugging Face implementation. The model is a RoBERTa-base encoder~\citep{liu2019roberta} with 12 attention heads that was pretrained on tweets and fine-tuned for three-class sentiment analysis in the TimeLMs project~\citep{loureiro2022timelms}. We run it with \texttt{attn\_implementation="eager"}, in fp32 under \texttt{torch.inference\_mode()}\footnote{\url{https://huggingface.co/cardiffnlp/twitter-roberta-base-sentiment-latest} (revision \texttt{3216a57f})}. The inputs are the first 2048 tweets of the sentiment test split of the TweetEval benchmark~\citep{barbieri2020tweeteval} (revision \texttt{b3a375ba}), tokenized once with padding to $L = 512$ and reused unchanged in every condition. Environment: the machine of Table~\ref{tab:env} with macOS 27.0, PyTorch 2.14.0 and Transformers 5.17.0.

\paragraph{Reference.} In inference mode a Transformer has no interaction between samples, so running the batch in one call must give the same outputs as running it in pieces, $f(\operatorname{concat}(X_1,\ldots,X_k)) = \operatorname{concat}(f(X_1),\ldots,f(X_k))$. We use this metamorphic relation as the reference: the same model on MPS with chunks of 256 texts, whose attention scores have $8.1\times10^{8}$ elements. We checked the reference against CPU fp32 on 1000 texts, including the first, middle and last 100, texts 1344--1384 around the predicted boundary, and the texts with the smallest margin between the top two logits. The largest relative logit difference was $6\times10^{-5}$, and no prediction changed.

\paragraph{Batch sizes.} Attention scores have $B \cdot 12 \cdot 512^2$ elements, so $B = 1365$ is just below $\pow{32}$ and $B = 1366$ is just above it. $B = 2048$ gives $1.5 \times \pow{32}$. At $L = 512$ the feed-forward activations stay below $\pow{32}$ at all three sizes, so only the attention products cross the boundary. We call a text corrupted when its relative logit error exceeds $10^{-3}$.

\begin{table}[htb]
\centering
\caption{Sentiment classification on MPS (PyTorch 2.14.0, macOS 27.0) with the whole batch in one forward pass, compared with chunked execution. ``Scores'' is the number of attention-score elements minus $\pow{32}$. Accuracy is on the three-class TweetEval labels; the value in parentheses is the reference.}
\label{tab:case}
\small
\begin{tabular}{rlrrrrl}
\toprule
$B$ & attention & scores $-\,\pow{32}$ & corrupted texts & changed predictions & max.\ rel.\ error & accuracy \\
\midrule
1365 & eager & $-1{,}048{,}576$ & 0 & 0 & 0 & 0.708 (0.708) \\
1366 & eager & $+2{,}097{,}152$ & 1 & 1 & 0.78 & 0.707 (0.708) \\
1366 & SDPA & $+2{,}097{,}152$ & 0 & 0 & $1.9\times10^{-5}$ & 0.708 (0.708) \\
2048 & eager & $+\pow{31}$ & 683 (33.4\%) & 386 (18.9\%) & 5.16 & 0.639 (0.713) \\
2048 & SDPA & $+\pow{31}$ & 0 & 0 & $1.9\times10^{-5}$ & 0.713 (0.713) \\
\bottomrule
\end{tabular}
\end{table}

\begin{figure}[htb]
\centering
\includegraphics[width=0.8\linewidth]{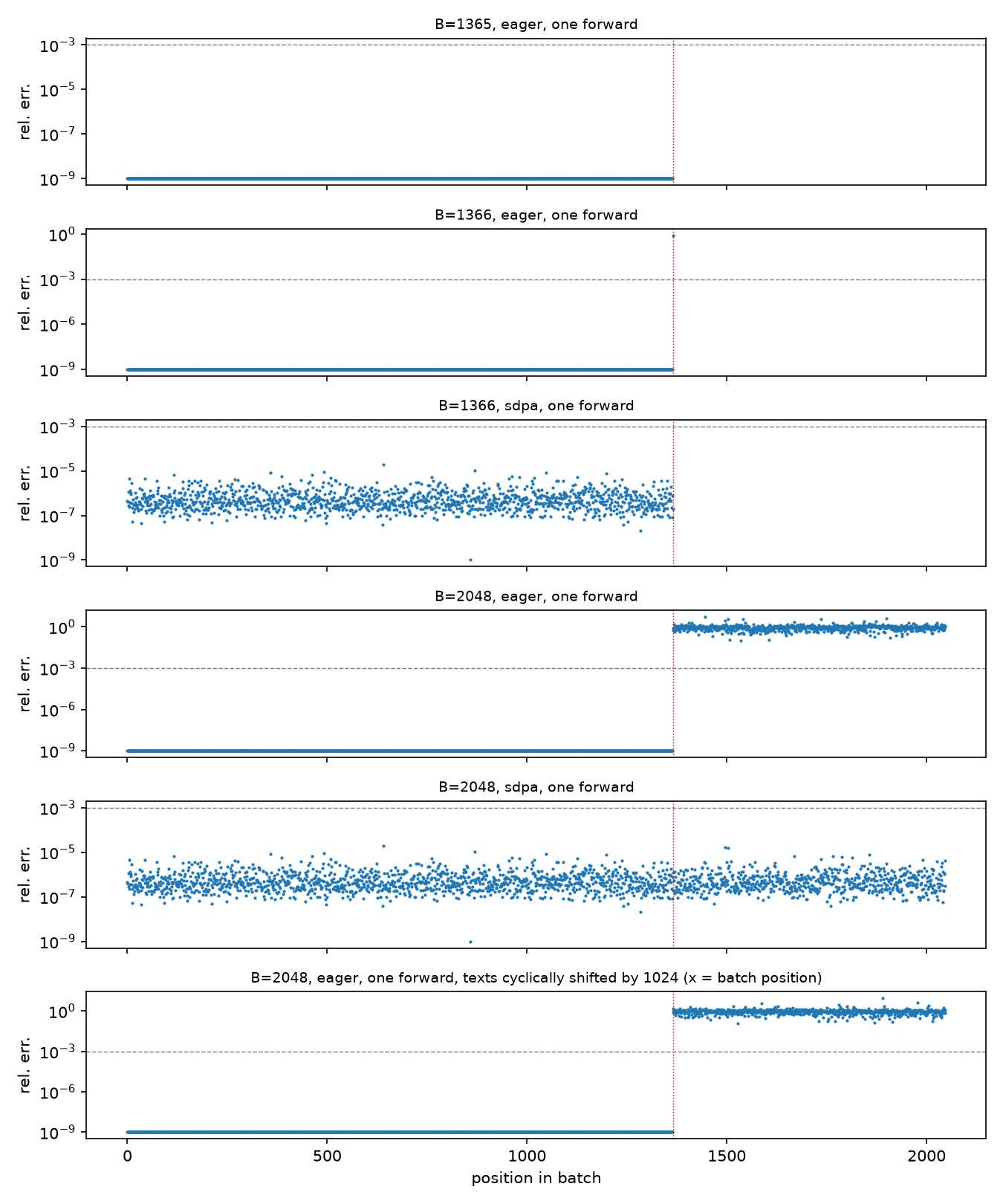}
\caption{Relative logit error of each text against the chunked reference, by position in the batch. Exact agreement is drawn at $10^{-9}$. The dotted red line marks text 1365, the first text predicted to cross $\pow{32}$ score elements; the dashed gray line is the $10^{-3}$ threshold. The last panel runs the same 2048 texts cyclically shifted by 1024.}
\label{fig:case}
\end{figure}

\paragraph{Results.} Table~\ref{tab:case} and Figure~\ref{fig:case} show the outcome. Every run finished normally, with no error, warning, NaN or Inf. At $B = 1365$ the single call matches the reference bit for bit. At $B = 1366$ exactly one text, the last one, is corrupted and its prediction changes. At $B = 2048$ texts 1365--2047 are corrupted, 386 predictions (18.9\%) change, and accuracy drops from 0.713 to 0.639. Every text before position 1365 still matches the reference bit for bit. Fused SDPA on the same batches agrees with the reference to within $1.9\times10^{-5}$.

\paragraph{Effect on predictions.} At $B = 2048$ the two groups of positions do not overlap: all 1365 texts before position 1365 have a relative error of exactly zero, and all 683 texts from position 1365 onward exceed the threshold, with a minimum of 0.10 and a median of 0.91. The corrupted outputs are not spread over the classes. Of the 683 corrupted texts, 661 are predicted as \emph{neutral}, whereas the reference assigns 242, 297 and 144 texts to negative, neutral and positive. Accuracy on these 683 texts falls from 0.723 to 0.501, which is what a constant \emph{neutral} prediction would score, since 344 of them (0.504) carry that label. The reference is correct and the single call wrong on 266 texts, and the reverse holds on 114 (exact McNemar test, $p = 4.1\times10^{-15}$). The damage is therefore much larger per class than overall. Over all 2048 texts, accuracy drops by 7.4 points, while recall drops from 0.788 to 0.529 for negative and from 0.740 to 0.504 for positive, and rises from 0.656 to 0.758 for neutral; on the 683 corrupted texts, recall for negative and positive is 0.010 and 0.054. Studies of nondeterminism in training observe the same pattern, with per-class and subgroup metrics varying far more than overall accuracy~\citep{pham2020variance,zhuang2021randomness}, although the cause there is rounding-level variation amplified by training and here it is an order-one error in a single forward pass. Because the corrupted outputs collapse onto one class, the visible loss depends on the label distribution, and a data set dominated by that class could hide it.

\paragraph{Where the corruption starts.} One head's score matrix has $512^2 = \pow{18}$ elements, so $\pow{32}$ elements correspond to $\pow{14} = 16384 = 1365 \times 12 + 4$ head matrices. If positions past element $\pow{32}$ of the flattened scores are affected, corruption should begin at head 4 of text 1365. Comparing the intermediate tensors of the first attention layer with the reference, the scores $QK^\top$ and the softmax output agree bit for bit at every batch size, and the first corrupted (text, head) pair in the product $\mathrm{softmax}(QK^\top)V$ is (1365, 4), with heads 4--11 of text 1365 corrupted. Feeding the reference softmax output into that product alone reproduces the same corruption. This matches the layer-level result that \texttt{bmm} fails when an input exceeds $\pow{32}$ elements.

\paragraph{Position, not content.} We repeated $B = 2048$ with the texts cyclically shifted by 1024. The corrupted batch positions are again 1365--2047, and the corrupted texts (original indices 341--1023) do not overlap with those of the unshifted run. Whether a text is misclassified depends on where it sits in the batch, not on what it says.

\paragraph{A loud failure at $L = 256$.} We first tried $L = 256$. There the feed-forward activation $(B \cdot 256) \times 3072$ crosses $\pow{32}$ at the same $B = 5462$ as the scores, and the process aborted inside a Metal matrix-multiplication assertion (\texttt{destination [3072, 1398272] is too large for kernel}), while $B = 5461$ matched the reference exactly. Whether a given oversized operation fails loudly or silently therefore depends on which operation crosses the boundary first.

\subsection{Other operations and CUDA}
\label{sec:firstpass}

\paragraph{First-pass suite.} Before the sweep we ran a suite of 26 test cases on all 11 PyTorch releases, with one fixed shape per case and a comparison against the CPU at sampled batch indices. Table~\ref{tab:versions} shows the cases that involve matrix multiplication and attention. The \texttt{bmm} rows agree with the sweep, and the suite adds two observations. Eager attention is silently wrong from 2.5.1 onward, as expected from the \texttt{bmm} results, except on 2.13.0, where the process aborts. On 2.4.1 it is correct at this shape, which the \texttt{bmm} rows alone do not explain. Fused SDPA is silently wrong up to 2.12.1 when its scores exceed $\pow{32}$ elements and correct from 2.13.0. A dispatch trace shows that on MPS fused SDPA resolves to \texttt{aten::\_scaled\_dot\_product\_attention\_math\_for\_mps} in both 2.12.1 and 2.14.0, and not to the FlashAttention operators that PyTorch provides for CUDA.

\begin{table}[htb]
\centering
\caption{First-pass results on macOS 27.0 by PyTorch version. \WRONG: silently wrong (relative error against CPU above $10^{-2}$); ok: correct; err: raises; crash: process aborts. Output-side cases use $(B,256,64)\times(B,64,256)$ with $B = 65552$ or $68544$; the input-side case uses $(B,256,256)\times(B,256,64)$ with $B = 65552$; attention uses scores of shape $(5712, 12, 256, 256)$.}
\label{tab:versions}
\small
\setlength{\tabcolsep}{3pt}
\begin{tabular}{lccccccccccc}
\toprule
case & 2.4.1 & 2.5.1 & 2.6.0 & 2.7.1 & 2.8.0 & 2.9.1 & 2.10.0 & 2.11.0 & 2.12.1 & 2.13.0 & 2.14.0 \\
\midrule
bmm, output $>\pow{32}$, contiguous & \WRONG & ok & ok & ok & ok & ok & ok & ok & ok & ok & ok \\
bmm, output $>\pow{32}$, $b$ transposed & \WRONG & \WRONG & \WRONG & \WRONG & \WRONG & \WRONG & \WRONG & \WRONG & \WRONG & \WRONG & \WRONG \\
bmm, output $>\pow{32}$, $a$ transposed & \WRONG & \WRONG & \WRONG & \WRONG & \WRONG & \WRONG & \WRONG & \WRONG & \WRONG & \WRONG & \WRONG \\
bmm, input $>\pow{32}$, contiguous & \WRONG & \WRONG & \WRONG & \WRONG & \WRONG & \WRONG & \WRONG & \WRONG & \WRONG & \WRONG & \WRONG \\
eager attention & ok & \WRONG & \WRONG & \WRONG & \WRONG & \WRONG & \WRONG & \WRONG & \WRONG & crash & \WRONG \\
fused SDPA & \WRONG & \WRONG & \WRONG & \WRONG & \WRONG & \WRONG & \WRONG & \WRONG & \WRONG & ok & ok \\
\bottomrule
\end{tabular}
\end{table}

\paragraph{Regression in 2.14.0.} On 2.14.0, \texttt{torch.arange} with $N = \pow{32} + \pow{20}$ elements writes only the first $N \bmod \pow{32}$ elements and leaves the rest zero without an error. Versions 2.12.1 and 2.13.0 raise instead (\texttt{MPSGraph does not support tensor dims larger than INT\_MAX}). This is the only regression that we found in 2.14.0; for \texttt{bmm} the sweep shows no difference between 2.10.0 and 2.14.0 (Table~\ref{tab:sweepversions}).

\paragraph{CUDA as a control.} Correctness in this paper is always judged against float64 on the CPU. CUDA serves a different purpose: it shows whether a failure belongs to the operation at that size or to the MPS backend. The controls for the backward and the extension series are reported with those series (Sections~\ref{sec:extended} and~\ref{sec:backward}). The main sweep on the A100 (PyTorch 2.14.0 with CUDA 13.0) comprises 1494 runs in 18 chunks, 54 of them calibration runs, over the same six shapes, two dtypes, four layouts and 42 batch sizes, with 992 runs on random and 448 on index-encoded inputs outside calibration. All 1494 runs are correct: no run is wrong, raises, crashes or returns zeros. This includes the 288 runs in which a tensor has more than $\pow{32}$ elements, 160 of them compared over all elements, which are the configurations where MPS fails. The largest error over all runs is $1.8\times10^{-6}$ in fp32 and $4.8\times10^{-4}$ in fp16, well below the tolerance of every chunk ($6.0\times10^{-6}$ to $1.5\times10^{-5}$ in fp32, $3.8\times10^{-3}$ and $4.2\times10^{-3}$ in fp16) and of the MPS sweep. Of the 1314 runs that the two sweeps have in common, 203 have a different outcome, and in all of them MPS raises (159) or is silently wrong (44) where CUDA is correct. The number of runs is lower than on MPS (1596) only because there are no switch points to compare on both sides. Before the sweep, we had also run the first-pass \texttt{bmm} configurations on the same GPU model with PyTorch 2.11.0 and CUDA 12.8 (48 configurations, all correct, largest error $9.2\times10^{-7}$ in fp32 and $4.5\times10^{-4}$ in fp16 against float64 on the GPU, with spot checks against the CPU), together with eager attention and fused SDPA with scores of shape $(5712, 12, 256, 256)$, which are correct with errors below $10^{-6}$. The \texttt{bmm} failures are therefore specific to the MPS backend and are not inherent to the operation, the shapes or the memory layouts. CUDA is not free of problems at this boundary, however. On the same A100, \texttt{torch.arange} with $N = \pow{32} + \pow{20}$ int64 elements leaves every element from index $\pow{32}$ onward at zero, without an error, in all four versions we tried, 2.11.0 to 2.14.0 (\ghissue{197673}). A fix was proposed within a day of the report; its author attributes the failure to an index that is multiplied in 32 bits before it is widened, in a kernel that \texttt{arange} shares with \texttt{range}, \texttt{linspace} and \texttt{logspace} (pull request \ghpr{197713}).\footnote{Open at the time of writing. We did not examine the source ourselves.} MPS on 2.14.0 fails differently on the same call, writing only the first $N \bmod \pow{32}$ elements. Two consequences follow. A second GPU backend cannot serve as the oracle for a differential test at this scale, which is why we use the CPU in float64 and, for \texttt{arange}, the closed form $x_i = i$. And silent failures at $\pow{32}$ elements are not confined to one backend, which suggests that this size range is rarely exercised by existing tests.

\section{Mitigation}
\label{sec:mitigation}

\paragraph{One rule.} A result that is silently wrong is worse than an error, so the mitigation should turn the first into the second. The rules of Table~\ref{tab:rules} could in principle be used to let the safe cases through, but Table~\ref{tab:sweepversions} shows that they change between versions, and Table~\ref{tab:versions} shows that attention and fused SDPA fail on their own schedule. We therefore use a single rule that refers to neither the version nor the operation: \emph{do not create or read an MPS tensor with $\pow{32}$ or more elements}. Large batches are split into chunks before they reach the device, as in the chunked reference of Section~\ref{sec:case}. \texttt{bmm}, and attention with it, has no interaction between batches, so chunking does not change the result. The same rule covers training. In the backward series the gradients are wrong only where the upstream gradient, which has the size of the forward output, exceeds $\pow{32}$ elements (Section~\ref{sec:backward}); a forward pass that respects the rule never produces such a gradient, and gradients accumulated over chunks equal those of the whole batch when the loss is a sum over samples. The offset series (Section~\ref{sec:extended}) suggests that contiguous chunks could also be taken from a large tensor that is already on the device, but this was measured on one version only, and the rule does not rely on it.

\paragraph{Guard.} The released guard enforces the rule. It is a \texttt{TorchDispatchMode} that inspects the inputs and outputs of every aten operation on MPS and raises an exception when one of them has $\pow{32}$ or more elements. For \texttt{mm}, \texttt{bmm}, \texttt{addmm}, \texttt{baddbmm} and fused SDPA it estimates the size before the operation runs, from the operand shapes (for SDPA, from the shape of the scores), because some versions abort the process inside the operation. For \texttt{bmm} the message states how many batches per chunk keep every operand and the output below the limit. The guard is installed with one call and needs no change to the model code.

\paragraph{Evaluation against the sweep.} We applied the stopping rule to the recorded outcome of every run (Table~\ref{tab:guard}). Over all 11 versions, 481 of 5252 runs are silently wrong or return zeros, and the rule stops all of them. The threshold has to be inclusive: with ``more than $\pow{32}$'' the rule would pass 16 runs on 2.4.1 in which a view of exactly $\pow{32}$ elements returns zeros. In no version is a run below $\pow{32}$ elements silently wrong. The price is that the rule also stops 630 of the 4095 correct runs, namely those in which a tensor of $\pow{32}$ or more elements happens to be handled correctly by that version. The rule fares the same on the series of Sections~\ref{sec:extended} and~\ref{sec:backward}, where the size of a run includes the upstream gradient, the gradients and the storage behind an offset view: all 104 silent runs are stopped, among them the 20 runs of the backward series in which only the gradients are wrong, and 128 of 730 correct runs are stopped as well, 28 of them offset runs whose storage reaches $\pow{32}$ elements. Run on the first-pass suite, the guard itself leaves no silent failure and no crash on 2.11.0, 2.12.1 and 2.14.0, and on the backward case with contiguous operands at $B = 65537$ it raises before the forward product is computed.

\begin{table}[htb]
\centering
\caption{The stopping rule applied to the recorded outcomes of every series on machine A. ``Silent'' counts runs that are silently wrong or return zeros; for the backward series, in any of the three tensors.}
\label{tab:guard}
\small
\begin{tabular}{lrrrrr}
\toprule
& & & \multicolumn{2}{c}{silent runs stopped} & \\
\cmidrule(lr){4-5}
PyTorch & runs & silent runs & $n \ge \pow{32}$ & $n > \pow{32}$ & correct runs stopped \\
\midrule
2.4.1 & 360 & 56 & 56 & 40 & 40 of 304 \\
2.5.1--2.8.0 (each) & 374 & 42 & 42 & 42 & 48 of 282 \\
2.9.1 & 360 & 32 & 32 & 32 & 40 of 274 \\
2.10.0--2.13.0 (each) & 360 & 32 & 32 & 32 & 40 of 276 \\
2.14.0 (main sweep) & 1596 & 97 & 97 & 97 & 198 of 1285 \\
\midrule
all versions, main and reduced sweeps & 5252 & 481 & 481 & 465 & 630 of 4095 \\
\midrule
2.14.0, $(256, 256, 256)$ shape & 264 & 32 & 32 & 32 & 28 of 154 \\
2.14.0, far points & 124 & 32 & 32 & 32 & 40 of 52 \\
2.14.0, offset series & 104 & 0 & 0 & 0 & 28 of 104 \\
2.14.0, backward pass & 604 & 40 & 40 & 40 & 32 of 420 \\
\bottomrule
\end{tabular}
\end{table}

\paragraph{Overhead.} The guard adds Python work to every operation. On a small training loop (a two-layer network on $1024 \times 120$ inputs, 2000 steps) it increases the run time by a factor of 1.73, and on large attention forward passes (batch 64, 32 heads, length 1024, fp16) by a factor of 1.00. The cost is thus negligible in the workload with large tensors, where the guard is needed, and the guard can be left out in workloads whose tensors are small by construction.

\section{Conclusion}

\subsection{Key Findings}

\begin{itemize}[leftmargin=*,itemsep=2pt]
\item \textbf{Silent failures beyond $\pow{32}$ elements follow three rules.} \texttt{bmm} on the MPS backend returns wrong values without an error in all 11 PyTorch releases from 2.4.1 to 2.14.0. Three rules, applied in order, reproduce all 1376 outcomes with random inputs. When the output exceeds $\pow{32}$ elements, it alone decides the outcome: a transposed operand makes every batch wrong, equal to a computation that ignores its strides, and anything else is correct. Otherwise a view with $\pow{31}$ or more elements raises an exception, and a contiguous input above $\pow{32}$ elements makes exactly the batches beyond that point wrong, equal to a computation whose index wraps around at $\pow{32}$. A slightly larger problem can therefore turn an explicit error into a silent failure. The limit is on the number of elements and not on bytes, it persists near $2 \cdot \pow{32}$ elements, and index-encoded inputs confirm both misreadings element by element.
\item \textbf{Training is affected where inference is not.} With contiguous operands and an output just above $\pow{32}$ elements, the forward pass is correct and both gradients are silently wrong in the batches beyond element $\pow{32}$. No new rule is needed: the rules, applied to the two products that define the gradients, reproduce all 592 outcomes of the backward sweep, including the pass in which each exception is raised.
\item \textbf{The failures are deterministic and specific to the backend.} An M2 Max under macOS 26.6.2, the same machine under macOS 27.0 and an M2 Ultra under macOS 27.0 agree in all 6156 common runs of every series, down to the largest error, for correct and wrong results alike; neither the chip nor the step from macOS 26 to 27 changes any result. The same sweeps, including the backward pass, are correct on CUDA in all 2530 runs. CUDA is nevertheless not a safe oracle at this scale, because \texttt{torch.arange} is silently wrong above $\pow{32}$ elements there as well, in a different way.
\item \textbf{Real workloads are affected, and aggregate metrics hide it.} With eager attention, one oversized batch of a public sentiment classifier corrupts a third of the outputs. The corrupted predictions collapse onto one class, so overall accuracy falls by 7 points while recall for the other two classes falls to almost zero on the affected inputs. Which texts are affected depends on their position in the batch and not on their content.
\item \textbf{One rule is enough in practice.} Do not create or read an MPS tensor with $\pow{32}$ or more elements, and split large batches before they reach the device. Applied to the recorded outcomes of all 11 versions, this rule stops all 481 silent runs (wrong values or zeros), and no run below $\pow{32}$ elements is silently wrong. It also prevents the backward failures, because the forward output, which has the size of the upstream gradient, is stopped first. The threshold must be inclusive, because 2.4.1 returns zeros for a view of exactly $\pow{32}$ elements.
\end{itemize}

\subsection{Limitations}

The measurements come from two machines of one chip generation and from one operation, and several limits follow from that.

\paragraph{Environment.} The MPS results come from two machines of the M2 generation, with macOS 26.6.2 and 27.0. We did not run macOS 14 or 15. That the behavior begins with macOS 15 rests on the error message that PyTorch raised before that release, and not on our own measurements. Other chip generations are untested. Repeating the sweep on older macOS versions is left for future work.

\paragraph{Coverage.} The systematic sweep covers \texttt{bmm} only. Other operations are covered by the first-pass suite with one shape each, so for them we know that a failure exists, but not where it begins. The sweep uses fp32 and fp16 and does not include bfloat16. Beyond $\pow{32}$ elements it has points just above the boundary and near $2 \cdot \pow{32}$, and nothing in between except the case study at 1.5 times $\pow{32}$. The backward series, the $(256, 256, 256)$ shape, the far points and the offset series use PyTorch 2.14.0 only. The backward series tests the gradients of a single \texttt{bmm}, not a training run. The offset series does not combine a large offset with a transposed view. The sweep of earlier PyTorch versions uses a coarser grid, one seed for the full comparisons and the base shapes only.

\paragraph{Interpretation.} The rules in Table~\ref{tab:rules} summarize observed outcomes for one PyTorch version. They agree with every run, including the extension and backward series, which cover conditions outside the sweep from which the rules were derived, but they describe a finite set of configurations and may not hold outside it. That the rules also reproduce the backward series rests on an assumption about how the two gradient products receive their operands, which we did not observe. The CUDA control covers every series on one GPU model and one PyTorch version (2.14.0); earlier versions were run on CUDA only for the first-pass configurations (2.11.0).

\paragraph{Case study and guard.} The case study uses one model, one data set and one input length. The collapse of the corrupted predictions onto a single class is an observation on this example, and other models may fail differently. The guard was run on the first-pass suite, and its stopping rule was checked against the recorded outcomes of the sweep; these are the same operations that motivated it. Its overhead was measured on two workloads. We recommend building the chunks before they reach the device, as in the case study. Slicing contiguous chunks out of a device tensor that already has $\pow{32}$ or more elements was correct in the offset series, but only on one PyTorch version, and creating such a tensor is what the rule forbids.

\section*{Acknowledgments}
This study is supported by JSPS KAKENHI (Grant No. JP24K\allowbreak{}16472).

\bibliographystyle{apalike}
\bibliography{../mps-boundary}

\end{document}